\documentclass[11pt,letter]{article}
\usepackage{fullpage}
\usepackage{setspace}
\usepackage{parskip}
\usepackage{titlesec}
\usepackage{xcolor}
\usepackage{lineno}
\usepackage{times}

\PassOptionsToPackage{hyphens}{url}
\usepackage[colorlinks = true,
            linkcolor = blue,
            urlcolor  = blue,
            citecolor = blue,
            anchorcolor = blue]{hyperref}

\usepackage[round]{natbib}
\let\cite\citep

\renewenvironment{abstract}
  {{\bfseries\noindent{\large\abstractname}\par\nobreak}}
  {}

\titlespacing{\section}{0pt}{*3}{*1}
\titlespacing{\subsection}{0pt}{*2}{*0.5}
\titlespacing{\subsubsection}{0pt}{*1.5}{0pt}

\usepackage{authblk}

\makeatother

\usepackage{graphicx}
\usepackage[space]{grffile}
\usepackage{latexsym}
\usepackage{textcomp}
\usepackage{longtable}
\usepackage{multirow,booktabs}
\usepackage{amsfonts,amsmath,amssymb}
\providecommand\citet{\cite}
\providecommand\citep{\cite}

\newif\iflatexml\latexmlfalse

\DeclareGraphicsExtensions{.pdf,.PDF,.png,.PNG,.jpg,.JPG,.jpeg,.JPEG}

\usepackage[utf8]{inputenc}
\usepackage[ngerman,english]{babel}

\begin{document}

\title{The arXiv of the future will not look like the arXiv}

\author[1]{\small Alberto Pepe}%
\author[2, 3, 1]{\small Matteo Cantiello}%
\author[1]{\small Josh Nicholson}%

\affil[1]{\footnotesize Authorea, 97 South 6th Street, 3rd Floor, Brooklyn, NY 11249, USA}%
\affil[2]{\footnotesize Center for Computational Astrophysics, Flatiron Institute, 162 Fifth Avenue, New York, NY 10010, USA}%
\affil[3]{\footnotesize Department of Astrophysical Sciences, Princeton University, Peyton Hall, Princeton, NJ 08544, USA}%

\vspace{-1em}

  \date{}

\begingroup
\let\center\flushleft
\let\endcenter\endflushleft
\maketitle
\endgroup

\begin{abstract}
The arXiv is the most popular preprint repository in the world. Since
its inception in 1991, the arXiv has allowed researchers to freely share
publication-ready articles prior to formal peer review. The growth and
the popularity~of the arXiv emerged as a result of~ new technologies
that made document creation and dissemination easy,~and cultural
practices where collaboration and data sharing were dominant. The arXiv
represents a unique place in the history of research communication and
the Web itself, however it has arguably changed very little since its
creation. ~Here we look at the strengths and weaknesses of arXiv in an
effort to identify what possible improvements can be made based on new
technologies not previously available. Based on this, we argue that a
modern arXiv might in fact not look at all like the arXiv of today. 

{\bf Disclaimer:} This article has originally been written and posted on~\href{https://www.authorea.com/}{Authorea}, a collaborative online platform for technical and data-driven documents. Authorea is being developed to respond to some of the concerns with current methodology raised in this very piece, and as such is suggested as a possible future alternative to existing preprint servers.  
\end{abstract}

\section{Introduction}

{\label{124571}}

\begin{figure}[h!]
\begin{center}
\includegraphics[width=1.00\columnwidth]{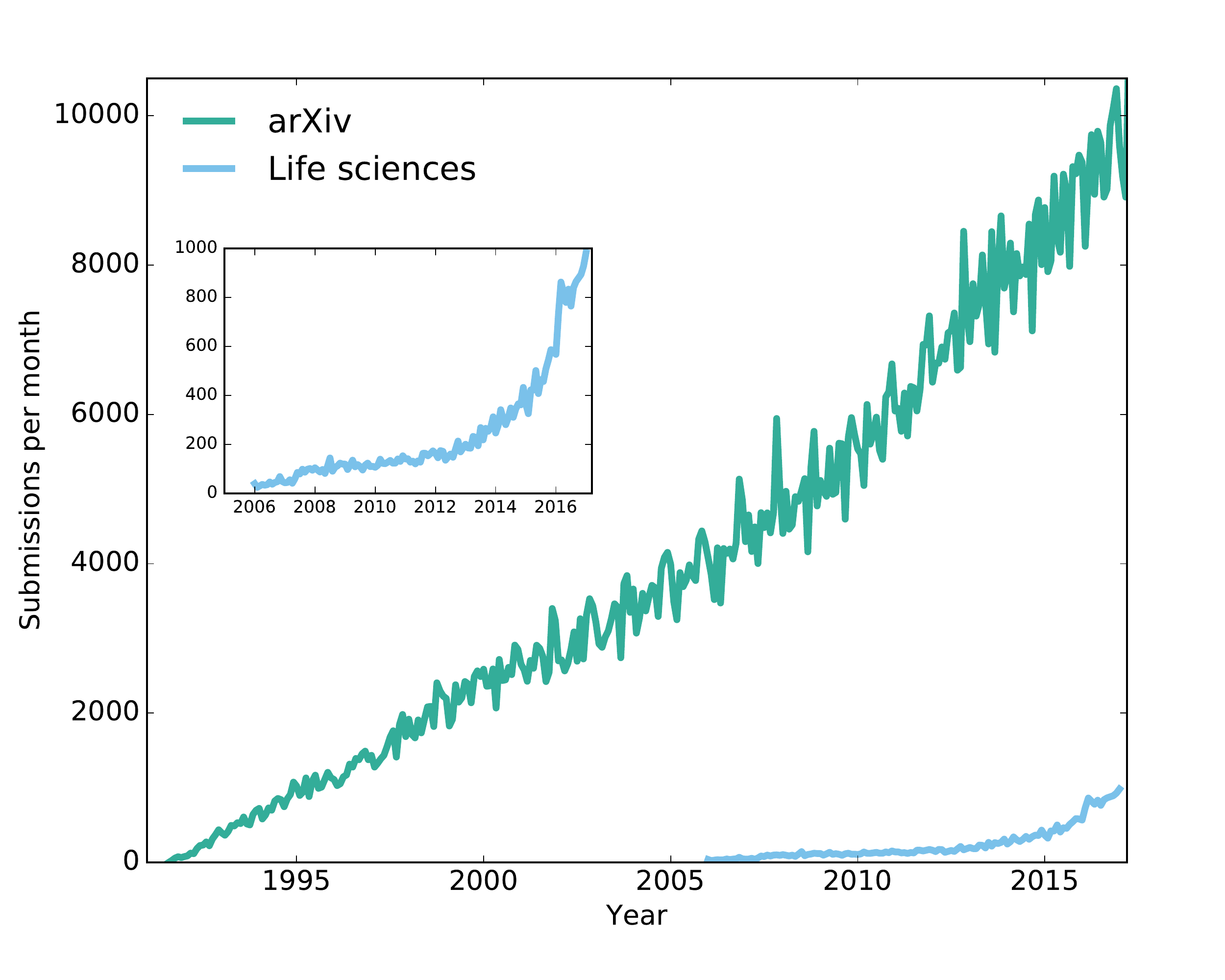}
\caption{{Volume of preprints posted in hard sciences (arXiv) and life sciences,
from 1991 to 2017. In this time window, the total number of arXiv (Life
sciences) preprints submitted was 1,263,265 (32,284). The inset shows
the recent, rapid growth of preprint submissions in the Life sciences
(including submissions to ``arXiv q-bio'', ``Nature Preceedings'',
``F1000Research'', ``PeerJ Preprints'', ``bioRxiv'', ``The Winnower'',
``preprints.org'' and ``Wellcome Open Research''). The data and the code
required to reproduce this figure is available in
the~\href{https://www.authorea.com/users/3/articles/173764-the-arxiv-of-the-future-will-not-look-like-the-arxiv}{web
version of this article} by clicking on the data and code
icons.~Unfortunately, these research products could not be included in the PDF version of the article due to the inherent limitations of this format. Data
Source:~\href{https://arxiv.org/stats/monthly_submissions}{arXiv}~and~\href{http://asapbio.org/preprint-info/biology-preprints-over-time}{Pre-Pubmed}.
{\label{104668}}%
}}
\end{center}
\end{figure}

The arXiv, pronounced ``archive'', is the~most popular preprint
repository in the world. ~Started in 1991 by physicist Paul Ginsparg,
the arXiv allows researchers to freely share publication-ready articles
prior to formal peer review and publication. Today, the arXiv publishes
over 10,000 articles each month from high-energy physics, computer
science, quantitative biology, statistics, quantitative finance, and
others (see Fig~{\ref{104668}}). The early success of
arXiv stems from the introduction of new technological advances paired
to a well-developed culture of collaboration and sharing. Indeed, before
the arXiv even existed, physicists were already physically sharing
recently finished manuscripts via mail, first, and by email, later. ~To
understand the success of the arXiv it is important to understand the
history of the arXiv. Below we highlight a brief history of technology,
services, and cultural norms that predate the arXiv and were integral to
its early and continued success. ~

\subsection{The history of the arXiv}

{\label{575743}}

Prior to the arXiv, ``the photocopy machine was a prime component of the
distribution system''~\cite{2011arXiv1108.2700G}~and ~preprints were only
exchanged to personal contacts and/or mailing lists ~\cite{Elizalde_2017}.
Institutional repositories, such as the SPIRES-HEP database (Stanford
Physics Information REtrieval System- High Energy Physics) at the
Stanford Linear Accelerator Center (SLAC) and the Document Server at
CERN only acted as bibliographic services, helping scientists to keep
track of publication information. But while SPIRES greatly improved the
flow of metadata, it was still hard to retrieve the full manuscript. A
new typesetting system would soon emerge and change this.

TeX, pronounced ``tech'', was developed by Donald Knuth in the late 70's
as a way for researchers to write and typeset articles programmatically.
Soon after the introduction of TeX, Leslie Lamport set a standard~for
TeX formatting, called LaTeX, which made it very easy for all
researchers to professionally typeset their own documents. This
system~made sharing papers easier and cheaper than ever before. Indeed,
many, if not most, researchers at the time relied upon secretaries or
typists to write their work, which then had to be photocopied in order
to be sent via mail to a handful of other researchers. TeX allowed
researchers to write their documents in a lightweight format that could
be emailed and then downloaded and compiled without the need for
physical mail.

Researchers began to exchange emails containing preprints, quickly
hitting their strict disk space allocation
limits ~\citep{Ginsparg_2011}. To address
this problem, an automated email server initially
called~\emph{xxx.lanl.gov} was set up on August 14th, 1991. This service
would allow researchers to automatically request preprints via email as
needed. It would soon become one of the world's first web servers and,
renamed~\emph{arXiv} in 1998, ~today still serves as one of the most
open and efficient forms of research communication in the world. ~

The arXiv was a leader in utilizing new technology when it was launched,
however it has arguably changed very little since its inception, despite
a wealth of new technologies now available. Here we look at the
strengths and weaknesses of the arXiv in an effort to identify what
possible improvements can be made based on new technologies and tools
and propose that a modern arXiv might in fact not look at all like the
arXiv of today, ~a development that will likely occur with or without
arXiv.

\par\null

\section{The strengths of the arXiv}

{\label{700903}}

The arXiv has since day one provided researchers with one of the easiest
and most powerful ways to disseminate their research. It is a free way
for authors to rapidly share findings directly with the research
community, and a free way for the public to access it. The arXiv is home
to some of the world's most important work, like the proof of
the~Poincar\selectlanguage{ngerman}é conjecture ~\citep{2002math.....11159P,2003math......3109P,2003math......7245P} and the discovery of the
Higgs Boson~\cite{1207.7235,1207.7214}. The free exchange of information has
been without equal in most other fields for nearly two decades until
very recently with the launch of numerous arXiv clones in new
disciplines (see Figure~{\ref{104668}}). The ease of
use and the utility of arXiv is both a function of the community it
serves--technically advanced researchers with a long-standing tradition
of sharing and collaboration--as well as the simplicity of the site.
Below we highlight key pieces of technology as well as cultural
influences that contributed to the success of arXiv. We then underline
in the next section how such pieces may be a limitation to new, and
better, practices.

\subsection{Typesetting with LaTeX}

{\label{707263}}

The vast majority of papers on arXiv are authored in LaTeX. LaTeX allows
researchers to easily typeset and share their documents. Such a solution
was available to all researchers at the outset, however it was only
adopted by the exact community it served, namely physicists and
mathematicians, who needed to write equation-intensive documents. Thus,
LaTeX was crucial to the early success of preprints and peer-to-peer
sharing. Today it continues to be used by physicists, mathematicians,
computer scientists, and others as it offers the best solution for
rendering complex mathematical notation.

\subsection{A tech-savvy community}

{\label{105418}}

The serendipitous arrival of new technology in a community that both
knew how to benefit from it and was willing to take advantage of it
(Physics), helped the arXiv to flourish from the very first day. Other
fields, like chemistry and biomedicine, while increasingly highly
collaborative in nature~\cite{Fanelli_2016}, may have lacked the early
knowledge and interest to write in LaTeX and set up and run email and
web servers, two necessary aspects to the foundation of the arXiv.
\par\null

\section{The weaknesses of the arXiv}

The immediate and sustained success of the arXiv since its inception is
due to its willingness to utilize new technology (LaTeX, email, web
servers) in a community naturally tech-savvy, collaborative, and open to
sharing practices. However, the arXiv has failed to improve and rethink
itself over time, to match the ever changing landscape of technology and
community practices in science. What is the single most important factor
that has prevented the arXiv to quickly innovate? We believe it is
LaTeX. The same technological advancement that has allowed the arXiv to
flourish, is also, incredibly, its most important shortcoming. Indeed,
the reliance of the arXiv on LaTeX is the source of all the weaknesses
listed below.

\subsection{Limitation to a single
community}

{\label{415532}}

Most researchers outside of physics, and consequently outside of the
arXiv world, write their manuscripts in Microsoft Word or other WYSIWYG
editors. Using LaTeX penetration rates in its most popular fields
(mathematics, statistics, physics, astronomy, computer science) it is
possible to estimate the total percentage of scholarly articles written
in LaTeX to be around 18\%~\cite{Pepea}. Not only does LaTeX have
a steep learning curve; its interface, language, and modus operandi are
foreign to anyone who does not program or to anyone who has only ever
used WYSIWYG word processors.

\subsection{\texorpdfstring{A printer-centric ``PDF
dump''}{A printer-centric PDF dump}}

{\label{969393}}

When you upload a LaTeX file, the arXiv compiles it and creates a PDF
document. This is a standard procedure. In academia, for decades
manuscripts have been exchanged and read in Postscript or PDF format.
PDFs are an efficient, portable format for printing manuscripts. But the
PDF is not a format fit for sharing, discussing, and reading on the web.
PDFs are (mostly) static, 2-dimensional and~\emph{non-actionable}
objects. It is not a stretch to say that a PDF is merely a digital
photograph of a piece of paper.

\subsection{Low discoverability}

{\label{697812}}

The research products hosted by the arXiv are PDFs. A title,~abstract,
and author list are provided by the authors upon submission as metadata,
which is posted alongside the PDF, and is rendered in HTML to aid
article discoverability. While search engines are getting better at text
mining PDFs, the chances that any current or future search engine will
meaningfully extract and interpret text from a dense 2-column paper are
low. Importantly, it is a futile exercise of reverse engineering. Why
are we locking content in a format that is not machine-readable?

\subsection{Data}

{\label{214609}}

Data sharing has become a fundamental practice across all scholarly
disciplines. Simply, if a published research paper is built on data, the
authors have to provide access to the minimal set of resources (data and
code) upon which their research is based. But sharing data in arXiv's
``LaTeX to PDF'' paradigm is not possible. A pilot to support data
deposit alongside papers which was run at the arXiv from 2010 to
2013~\cite{Mayernik_2012}, failed to gain traction. While the project had
to face an unexpected cut in government support, we believe that part of
its failure can be associated with the fact that the papers and the data
were deposited as separate entities. How do people share data today?
They use kludgy strategies. A growing trend in astronomy and physics,
for example, is to link the dataset in the published or preprinted
paper. This practice allows authors to make their data more visible and
get credit for it, as it is linked inside the papers, but recent work
shows that links rot quickly with time~\cite{Pepe_2014}.

\par\null

\section{How the arXiv would look like if we built it
today}

{\label{332132}}

A useful exercise when attempting to imagine the arXiv of the future is
to envision what it would look like if we could rebuild it today. We
would like to consider the weaknesses listed above as opportunities
rather than challenges, and in doing so, we offer here some ideas for a
better arXiv.

\subsection{Web-native and web-first}

{\label{285318}}

There is growing consensus in scholarly communication circles that
academic publishing needs to move ``beyond the PDF'' (see the manifesto
of~\href{https://www.force11.org/}{Force 11}), and we strongly believe
that the paper of the future will be web-native~\cite{Goodman}. As
such, the arXiv of the future will have to enable creation and/or
ingestion of papers in HTML format. Moving scholarly papers to HTML is
the first step towards paving the way for the scholarly repository of
the future. The paper you are currently reading, whether you are reading
a PDF or HTML version of it, is web-first. An open, web version can be
found~\href{https://www.authorea.com/173764}{here}.~\textbf{The arXiv of
the future will host web-native manuscripts}.

\subsection{Multi-format and
format-neutral}

{\label{287186}}

The ArXiv relies heavily on LaTeX. The article you are reading was
authored on Authorea by three authors using a combination of LaTeX and
Rich Text. LaTeX was just a format used to insert mathematical notation,
equations, tables, but not to typeset and format the entire
manuscript.~LaTeX can be a time consuming way to typeset
manuscripts~\cite{brischoux2009don}, most importantly it also locks the
document into a format which doesn't allow for the flexibility offered
by modern technologies (e.g.~semantic parsing and embedding into a
knowledge network that facilitates discoverability, hence
impact)~.~\textbf{The arXiv of the future is format-neutral and
separates format from content.}

\subsection{Digital object
identifiers}

{\label{200799}}

A digital object identifier (DOI) is a persistent identifier used in
scholarly publishing to identify and link to a piece of work.~ DOIs are
considered by numerous journals to be mandatory for citation and can be
assigned to datasets, preprints, research articles, websites, and other
scholarly works. Since the practice of preprinting is quickly on the
rise across all disciplines \cite{berg2016preprints}, and since funding bodies
are finally realizing the importance of preprinting (here's
a~\href{https://grants.nih.gov/grants/guide/notice-files/NOT-OD-17-050.html}{recent
example by the NIH}), it is crucial that preprints get identified by a
reliable standard: the DOI. The article you are reading was written on
Authorea and it was preprinted with DOI
(\url{https://dx.doi.org/10.22541/au.149693987.70506124}). \textbf{The
arXiv of the future is a database of preprints identified by DOI.}

\subsection{Built for Open Data and Open
Research}

{\label{587335}}

The repository of the future is more than a collection of PDFs with text
and images. The repository of the future is a database of papers that
integrate data, code and all the resources needed to reproduce
scientific results. The only way to solve the ongoing reproducibility
crisis is by making papers data-driven. The paper you are reading has
one figure and we have made the data ``behind the figure'' available to
all readers. If you read
the~\href{https://www.authorea.com/users/3/articles/173764-the-arxiv-of-the-future-will-not-look-like-the-arxiv}{online
version of the article}, you will be able to click on the Data flag
associated with Fig.~{\ref{104668}} and visualize,
download, and peruse the data presented in the chart, as well as the
code (in the form of a Jupyter Notebook) which we wrote to analyze and
visualize such chart.~\textbf{The arXiv of the future will host data and
code alongside papers}.

\subsection{Comments and open peer
review}

{\label{107257}}

The arXiv does not currently allow comments by its readers and authors.
The idea is that the arXiv is not peer-reviewed - peer-review happens
elsewhere, at the journal level. Thus, a comment and reviewing system is
hard to maintain and run, and not useful. Yet, preprints offer an
unprecedented opportunity to first, open up, and second, increase the
quantity of reviews and comments that manuscripts go through. We do not
advocate to replace traditional peer-review, but to complement it with
open review of preprints. We believe that (1) more scholars should
participate in the peer review of an article, and (2) peer review should
be done in the open, so that the review itself becomes a crucial
component of the published (or pre-printed) research. This seems to be
not only natural, but also necessary given the ever increasing number of
publications and  average number of authors per paper, which
renders the current peer-review paradigm unsustainable. The authors of
this document welcome public comments and ideas from its readers, at the
\href{https://www.authorea.com/173764}{online version of this article}.
\textbf{The arXiv of the future will allow open comments and reviews in
addition to traditional peer review.}~

\subsection{Alternative metrics}

{\label{504300}}

In scholarship, currently the only reputable metric to assess the impact
of a research paper is citations (or any other metric built around
citations). The arXiv does not publish information about alternative
metrics (alternative to citations), e.g. how many times a paper has been
downloaded, tweeted, or blogged. One important, yet dubious, case
against publishing these altmetrics is that they can be easily gamed.
And if these metrics gain traction and become a reputable system to
determine the standing of a researcher, then we are confronted with an
easily-gamed system. We believe that these metrics provide important
value in assessing the impact of research work, in addition, not as a
replacement, to traditional metrics.~Importantly, it has been shown that
there is a very strong correlation between how many times a paper is
downloaded and tweeted and how many times it is subsequently
cited~\cite{Haque_2009,Haque_2010,Shuai_2012}.~\textbf{The arXiv of the future will be
transparent and it will publish information about alternative metrics
that may determine the true impact of a research paper}.~

\subsection{Discoverable, structured, semantic,
machine-readable}

{\label{272554}}

One final but very important advantage that is tightly linked to having
a repository of natively web-based - rather than PDF - papers is
discoverability. The entire full text of papers - not just the title and
abstract - will be indexed by search engines and scholarly repositories,
boosting content visibility. Moreover, web-based articles have a
well-defined semantic structure, making them fully machine-readable
objects.~\textbf{The arXiv of the future will rethink papers as APIs
that access semantically structured content}.

\par\null

\section{Conclusion}

{\label{349165}}

{We have looked at the history of the arXiv, identifying a number of
possible reasons that determined its success as the most popular online
preprint repository. We argue that one of the reasons why the arXiv
flourished is because it catered to~}technically savvy researchers with
a long standing tradition of sharing and collaboration. ~The simplicity
of the site and its LaTeX-centric submission process, secured rapid
growth within communities~that were already used to taking full control
of the typesetting process and needed to write equation-intensive
documents.

{We argued that while the arXiv was quick to adopt technology early on,
it has changed very little since its launch. This unwillingness or
inability to foster new technology and practices are an impediment to
better research communication practices currently available.}

We suggest that the arXiv of the future will be web-native and
web-first, multi-format and format-neutral so as to include the whole
research community. In order to foster transparency and reproducibility,
it will be built for open data and open research, also allowing for
commenting and open peer review. The arXiv of the future will be a
database of preprints identified by a Digital Object Identifier, with a
well-defined semantic structure that will make them fully
machine-readable and easily discoverable. It will be transparent and
publish all the information about alternative metrics that may determine
the true impact of the research it hosts. ~

{We believe that should the arXiv continue to remain stagnant, it will
be eclipsed by other services, just like the arXiv itself did to its
predecessors. We encourage researchers to demand more out of the
platform and believe that in the era of the web, sharing research via
PDF must inevitably come to an end. Let us embrace new technologies and
practices, just like the arXiv did nearly 30 years ago so that we might
create a better way to share research.}

\par\null

{\textbf{Acknowledgments}}

{The authors would like to thank Paul Ginsparg, Konrad Hinsen, Michael
Kurtz ~and Deyan Ginev for their very insightful criticism and comments,
which helped to substantially improve the manuscript. Their
contributions can be found in the comment threads on
the~\href{https://www.authorea.com/173764}{online version of the
article}.~}

\selectlanguage{english}
\bibliographystyle{plainnat}
\bibliography{biblio.bib%
}

\end{document}